\documentclass[12pt]{article}%
\usepackage{amsmath}
\usepackage{amsfonts}
\usepackage{amssymb}
\usepackage{graphicx}%
\providecommand{\U}[1]{\protect\rule{.1in}{.1in}}

\begin{document}

\title{Severe discrepancies between experiment and theory in the superconducting
proximity effect}
\author{M.Zhang, G.Tateishi and G.Bergmann\\Department of Physics\\University of Southern California\\Los Angeles, California 90089-0484\\e-mail: bergmann@usc.edu\\D:%
$\backslash$%
0aa%
$\backslash$%
Aa\_tex%
$\backslash$%
A\_paper%
$\backslash$%
Arb\_\_%
$\backslash$%
B154}
\date{\today }
\maketitle

\begin{abstract}
The superconducting proximity effect is investigated for SN double layers in a
regime where the resulting transition temperature $T_{c}$ does not depend on
the mean free paths of the films and, within limits, not on the transparency
of the interface. This regime includes the thin film limit and the normalized
initial slope $S_{sn}=$ $(d_{s}/T_{s})|dT_{c}/dd_{n}|.$ The experimental
results for $T_{c}$ are compared with a numerical simulation which was
recently developed in our group. The results for the SN double layers can be
devided into three groups: (i) When N = Cu, Ag, Au, Mg a disagreement between
experiment and theory by a factor of the order of three is observed, (ii) When
N = Cd, Zn, Al the disagreement between experiment and theory is reduced to a
factor of about 1.5, (iii) When N = In, Sn a reasonably good agreement between
experiment and theory is observed.

\newpage

\end{abstract}

\section{Introduction}

The properties of a superconducting film or thin wire S are modified when they
are in contact with a normal metal N. This phenomenon was first observed in
the pioneering experiments by Meissner \cite{M71} who explored the properties
of superconducting wires covered with normal metals. It is generally called
the "superconducting proximity effect" (SPE). It was intensively studied in
the 1960's \cite{H31}, \cite{H26}, \cite{D36}, \cite{H27}, \cite{B133},
\cite{M45}. During the last decade it has experienced a renewed interest
theoretically \cite{A63}, \cite{B161}, \cite{A62}, \cite{A57}, \cite{Z7},
\cite{Z8}, \cite{V11}, \cite{M62}, \cite{S50}, \cite{S51}, \cite{T16},
\cite{N13}, \cite{Z9} as well as experimentally \cite{N9}, \cite{V14},
\cite{B144}, \cite{P31}, \cite{M62}, \cite{S51}, \cite{D40}, \cite{B134},
\cite{B135}, \cite{V16}. Recently the SPE has been extended to SN-multi layers
\cite{B163}, \cite{D41}.

A few years ago, our group \cite{B135} investigated the proximity effect
between Pb and several alkali metals. For a better analysis of these
measurements we developed a quantitative numerical method for the calculation
of the transition temperature of an SN double layer \cite{B149}. Our numerical
results show that, when a superconductor S is covered with a normal metal N,
that the initial slope $\frac{dT_{c}}{dd_{n}}$ is independent of the mean free
paths of the two metals and the transparency of the interface (if the
transmission is not dramatically changed). If one defines a normalized initial
slope $S_{sn}=\frac{d_{s}}{T_{s}}\left\vert \frac{dT_{c}}{dd_{n}}\right\vert $
then $S_{sn}$ is independent of the thickness $d_{s}$ of the superconductor up
to relatively large values of $d_{s}$. If the superconductor is very weak
coupling ($2\pi T_{s}<<\Theta_{D}$, $\Theta_{D}$=Debye temperature) then our
result for the initial slope converges towards the results for the thin film
Cooper limit \cite{C21} (see below).

When we compared our experimental initial slope $\frac{dT_{c}}{dd_{n}}%
|_{d_{n}=0}$ with our numerical calculation we observed that the experimental
results were considerably smaller than the theoretical predictions. Surprised
by the discrepancy we searched the literature and found early experiments from
the 1960s, particularly by Hilsch \cite{H31}, \cite{H26} and Minigerode
\cite{M45}, from which the normalized initial slope can be derived. These
measurements showed a similar disagreement in the initial slope with the
theory (see ref. \cite{B149}).

Since we were rather amazed by the discrepancy between our experiments and
theory in the SPE and also by the fact that this discrepancy had not been
detected previously, we decided to re-investigate the SPE. In this paper we
investigate the SPE in the range, where a minimum of experimental parameters
is needed to perform a quantitative comparison with the theory. We focus on
the normalized initial slope $S_{sn}$ of SN sandwiches and the transition
temperature of very thin NS sandwiches in the thin film limit.

\section{Experiment and Results}

We use thermal evaporation to condense the thin films onto a substrate at
liquid helium temperature. To obtain clean films all the evaporation sources
are surrounded with liquid N$_{2}$ and the vacuum in our system is better than
$10^{-11}$torr. We first evaporate 10 atomic layers of insulating Sb on a
helium cold quartz substrate. The Sb film acts as a fresh substrate and
insures that the following quench condensed films are flat and homogeneous.

In a series of experiments a film of the superconductor Pb is first condensed
onto the Sb substrate. Afterwards the Pb is covered in several step with an
increasing thickness of the normal metal Ag. The thickness of the films is
measured with a quartz oscillator. The accuracy of the thickness measurement
is about 15\%. After each evaporation the superconducting transition curve,
the magnetoresistance and the Hall effect of the double layer are measured.
Fig.1 shows a plot of $T_{c}$ versus the Ag thickness $d_{Ag}$ on top of a
$251A$ thick Pb film. This plot yields graphically the initial slope
$dT_{c}/dd_{n}|_{d_{n}=0}$ and the normalized initial slope $S_{sn}$%

\begin{equation}
S_{sn}=\frac{d_{s}}{T_{s}}\left\vert \frac{dT_{c}\left(  d_{n}=0\right)
}{dd_{n}}\right\vert \label{Ssn}%
\end{equation}
where $d_{Pb}=d_{s}$ is the thickness of the Pb films and $T_{s}=7.2K$ is the
transition temperature for Pb alone.%

\begin{align*}
&
{\includegraphics[
height=3.2461in,
width=3.8779in
]%
{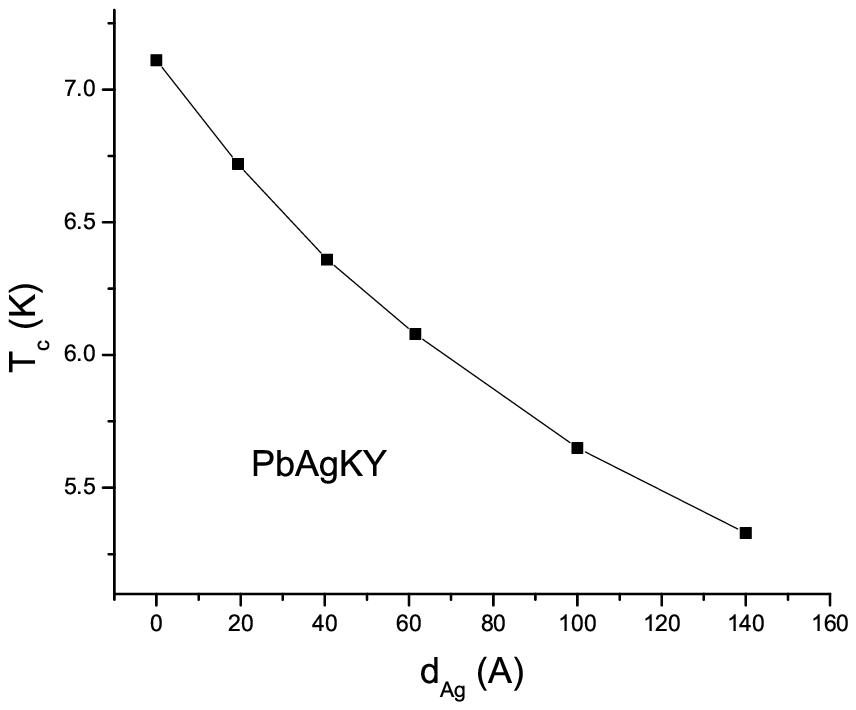}%
}%
\\
&
\begin{tabular}
[c]{l}%
Fig.1: $T_{c}$ versus $d_{Ag}$ for an PbAg double layer.
\end{tabular}
\end{align*}

This experiment is repeated for different thicknesses of the superconductor
Pb. In Fig.2 $S_{sn}$ is plotted versus the Pb thickness $d_{Pb}$. It is
essentially independent of the Pb thickness. This was the prediction of our
numerical results. The value of the normalized initial slope is $S_{PbAg}%
=0.66\pm0.05$.%

\[%
\begin{array}
[c]{c}%
{\includegraphics[
height=3.0369in,
width=4.0174in
]%
{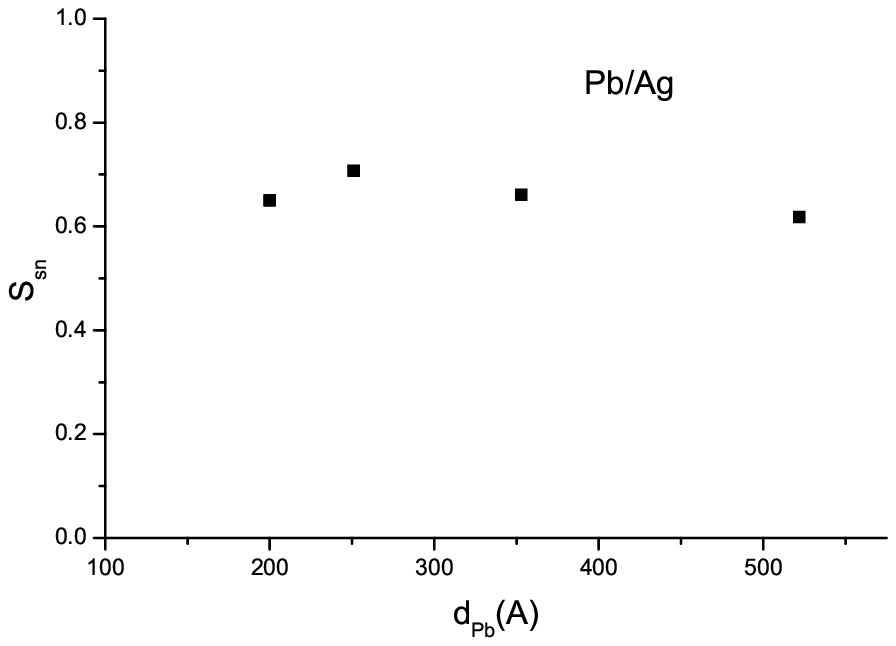}%
}%
\\%
\begin{array}
[c]{c}%
\begin{array}
[c]{c}%
\text{Fig.2: The normalized initial slope }S_{sn}\text{ for the Pb/Ag }\\
\text{sandwiches as a function of the Pb thickness}%
\end{array}
\end{array}
\end{array}
\]

\[
\]

In the next experimental series we investigate NS double layers in the thin
film limit and express the results in terms of the normalized initial slope
$S_{sn}$. As an example a thin Ag film ($d_{Ag}=41.0$A) is condensed onto the
insulating Sb substrate\ and covered in several steps with increasing
thickness of Pb. Fig.3a shows the results. The inverse $T_{c}$-reduction
$1/\Delta T_{c}=1/\left(  T_{s}-T_{c}\right)  $ is plotted as a function of
the Pb thickness. From these plots we extract the value of the normalized
initial slope. For the AgPb double layer the value is $S_{PbAg}=0.66.$ This is
in excellent agreement with the results from the first experimental series.

$%
\begin{array}
[c]{cc}%
{\includegraphics[
height=2.7231in,
width=3.1415in
]%
{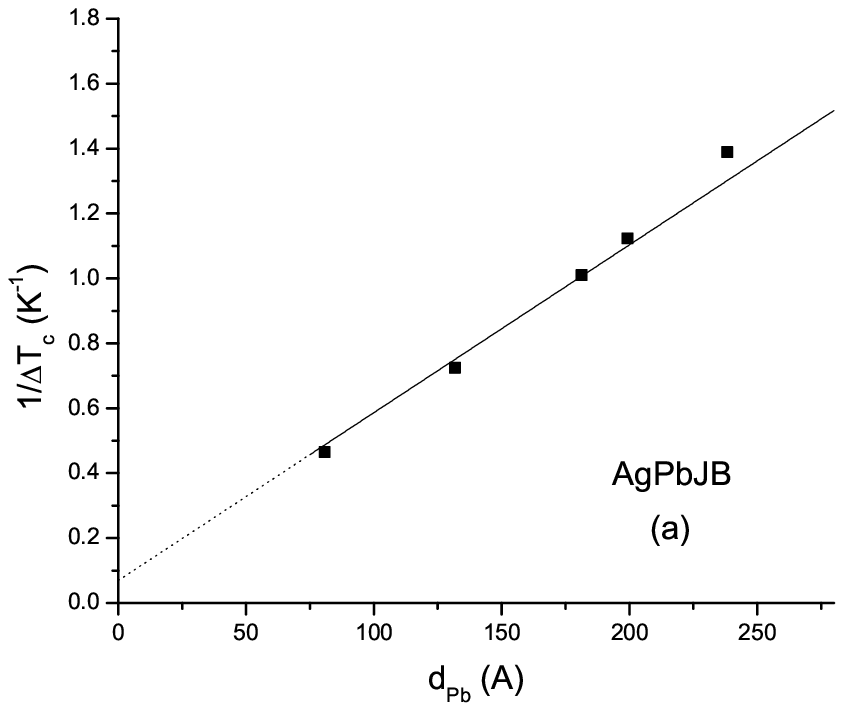}%
}%
&
{\includegraphics[
height=2.714in,
width=3.1996in
]%
{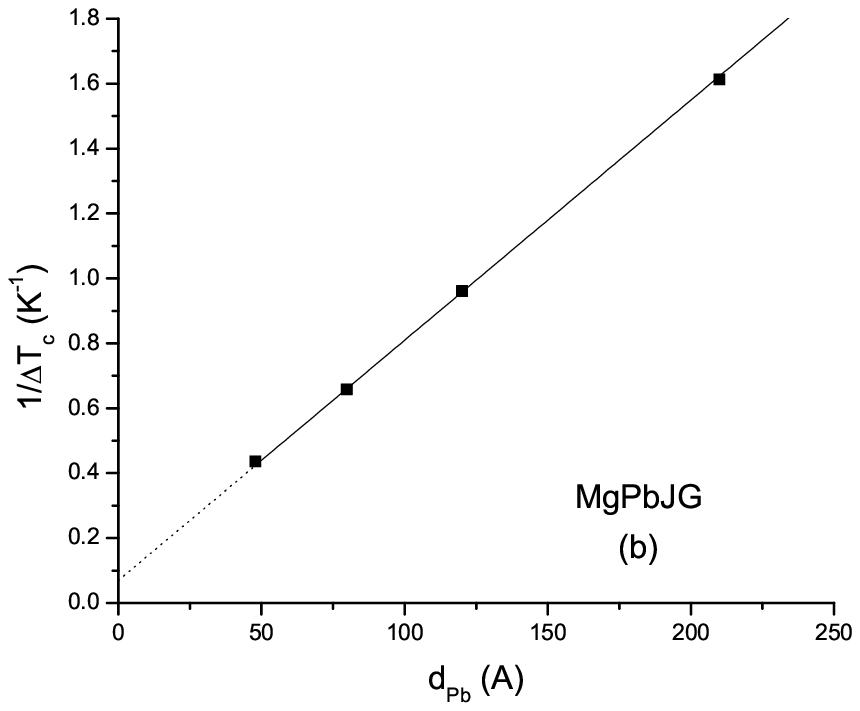}%
}%
\\
& \\%
{\includegraphics[
height=2.8269in,
width=3.32in
]%
{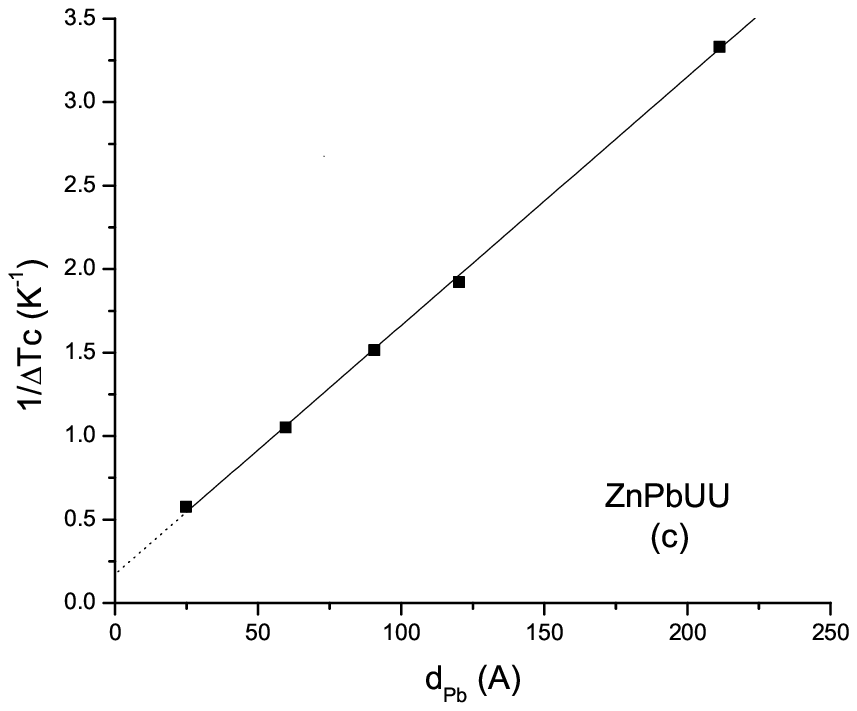}%
}%
&
{\includegraphics[
height=2.8244in,
width=3.2553in
]%
{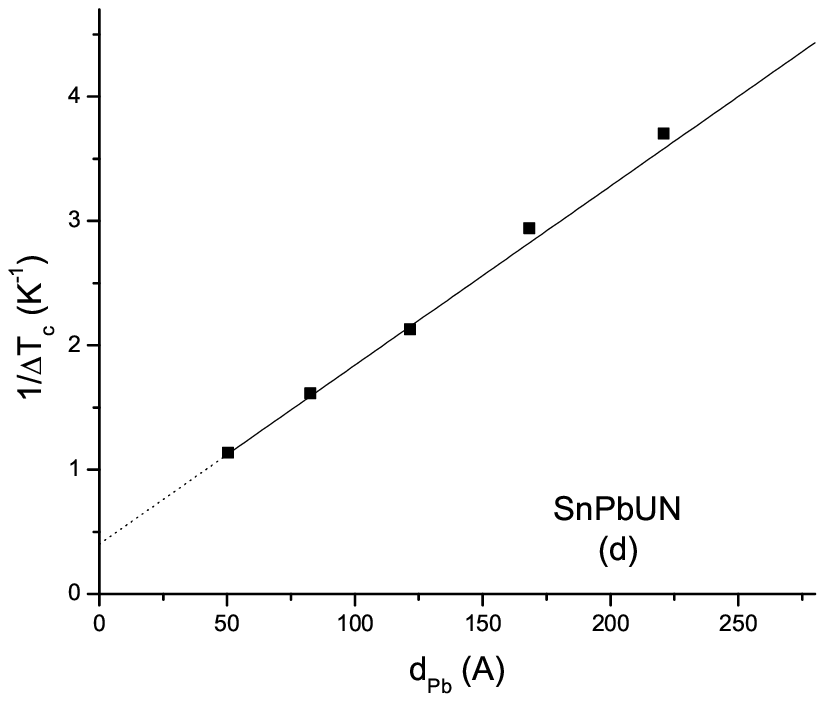}%
}%
\end{array}
$

$%
\begin{tabular}
[c]{l}%
$\text{Figure 3. The inverse }T_{c}\text{ reduction }1/\left(  T_{s}%
-T_{c}\right)  \text{ versus the Pb thickness }d_{Pb}\text{ }$\\
for$\text{ double layers }$of N/Pb where N stands for the metals Ag, Mg, Zn
and Sn.
\end{tabular}
$%

\[
\]

We use this much more efficient procedure to investigate sandwiches of Pb with
the normal metals Ag, Au, Cu, Mg. Furthermore we included also as "normal
metals" superconductors with a transition temperatures $T_{n}$ which lie below
the value of $T_{c}$ for Pb. These metals are Zn, Cd, Al, In and Sn. Fig.3b-d
shows some of the results for Mg, Zn and Sn.

In table I the experimental data are collected. The first column gives the
experimental code; which includes the symbols for the normal metal and the
superconductor. The following columns give the thickness of the normal metal
(or the superconductor with the lower $T_{c}$), the transition temperature
$T_{n}$ of N if N is a superconductor, the ratio of the experimental (i.e.,
phonon enhanced) density of states and the experimental normalized initial
slope $S_{\text{exp}}.$ The estimated error of $S_{\text{exp}}$ is about 15\%.
The density of states are taken from Kittel's book \cite{K53}. The theoretical
prediction of $\ $the normalized initial slope by the Werthamer theory
$S_{\text{Wh}}$ and our numerical result $S_{\text{sim }}$will be discussed
below. The transition temperatures for quench condensed Cd and Zn are taken
from ref. \cite{H33} and \cite{B162}, the transition temperatures of quench
condensed Al, In and Sn are taken from \cite{B32} and \cite{B6}.%

\begin{align*}
&
\begin{tabular}
[c]{|l|l|l|l|l|l|l|l|l|}\hline
\textbf{exp.} & $\mathbf{d}_{n}$(A) & $\mathbf{T}_{n}$(K) & $\mathbf{N}%
_{n}^{\ast}\mathbf{/N}_{s}^{\ast}$ & $\mathbf{S}_{\exp}$ & $\mathbf{S}%
_{\text{Wh.}}$ & $\mathbf{S}_{\text{sim.}}$ & $\mathbf{S}_{\exp}%
/\mathbf{S}_{\text{sim.}}$ & $\mathbf{S}_{\text{sim.}}/\left(  \mathbf{N}%
_{n}^{\ast}\mathbf{/N}_{s}^{\ast}\right)  $\\\hline
AgPbJB & 41.2 & 0 & 0.387 & 0.66 & 0.95 & 1.74 & 0.38 & 4.\thinspace5\\\hline
CuPbJE & 32.9 & 0 & 0.603 & 0.98 & 1.49 & 2.67 & 0.37 & 4.\thinspace
\allowbreak43\\\hline
AuPbJD & 29.5 & 0 & 0.442 & 0.62 & 1.09 & 1.98 & 0.31 & 4.\thinspace
\allowbreak48\\\hline
MgPbJG & 28.5 & 0 & 0.572 & 0.48 & 1.41 & 2.54 & 0.19 & 4.\thinspace
\allowbreak44\\\hline
CdPbJJ & 31.4 & 0.80 & 0.329 & 0.39 & 0.40 & 0.528 & 0.74 & 1.\thinspace
\allowbreak60\\\hline
ZnPbUU & 26.0 & 1.39 & 0.430 & 0.33 & 0.44 & 0.54 & 0.61 & 1.\thinspace
\allowbreak26\\\hline
AlPbUT & 21.5 & 2.28 & 0.833 & 0.44 & 0.68 & 0.76 & 0.58 & 0.91\\\hline
InPbUP & 32.1 & 4.1 & 0.663 & 0.30 & 0.31 & 0.354 & 0.85 & 0.53\\\hline
SnPbUN & 33.0 & 4.7 & 0.664 & 0.29 & 0.25 & 0.283 & 1.02 & 0.43\\\hline
&  &  &  &  &  &  &  & \\\hline
\end{tabular}
\\
&
\begin{tabular}
[c]{l}%
$\text{Table I: The experimental normalized initial slope }S_{\text{exp}%
}\text{ }$are compared with\\
the theoretical predictions by Werthamer's theory $S_{\text{Wh}}$ and the
author's\\
numerical calculations $S_{\text{sim}}$. The first four columns give the
experimental\\
code (containing the symbols of the normal conductor N and superconductor
S),\\
the thickness of N, the transition temperature of N (if superconducting) and\\
the ratio of the experimental density of states $\mathbf{N}_{n}^{\ast
}\mathbf{/N}_{s}^{\ast}$ (which includes\\
electron-phonon enhancement). The second to last column gives the ratio\\
between the experimental $S_{\text{exp}}$ and the numerical results
$S_{\text{sim}}$.
\end{tabular}
\end{align*}

\section{Discussion}

Let us consider a double layer with the superconducting transition temperature
$T_{c}$. This transition temperature defines a characteristic time
$\tau_{T_{c}}=\tau_{c}=\hbar/\left(  \pi k_{B}T_{c}\right)  $. The
superconducting coherence length is the distance that an electron propagates
during this time $\tau_{c}$. In a clean metal this is equal to $\xi_{0}%
=v_{F}\tau_{c}$ and in a dirty metal one has $\xi=\sqrt{D\tau_{c}}$ where $D$
is the diffusion constant of the dirty metal.

When, during the time $\tau_{c},$ an arbitrary conduction electron propagates
through the whole thickness range of the double layer with roughly equal
probability then the superconducting properties of the system are averaged
over both films. In this case the mean free paths of the individual metal have
no bearing on the superconducting transition temperature $T_{c}.$ (However,
the mean free paths determine whether the system is in this limit). A
well-known example is the thin film or \textquotedblright
Cooper\textquotedblright\ limit of a double layer when both films are much
thinner than their coherence lengths and the \textquotedblright transmission
time\textquotedblright\ through the interface is much shorter than $\tau_{c.}$
In the limit that $T_{c}\ $is much smaller than the Debye temperature,
$T_{c}<<\Theta_{D},$ the transition temperature of such a double layer is
given by the BCS-Cooper formula
\begin{equation}
T_{c}=1.14\Theta_{D}\exp\left(  -\frac{1}{\left(  NV\right)  _{ef}}\right)
\label{Tc_Cp}%
\end{equation}
where the effective BCS interaction $\left(  NV\right)  _{ef}$ of the double
layer is
\[
\left(  NV\right)  _{ef}=\frac{d_{s}N_{s}\left(  NV\right)  _{s}+d_{n}%
N_{n}\left(  NV\right)  _{n}}{d_{s}N_{s}+d_{n}N_{n}}%
\]
and $\left(  NV\right)  _{s,n}$ are the BCS interactions in the super- and
normal conductor.

Using the BCS-Cooper formula for an SN double layer yields for the normalized
initial slope
\begin{equation}
S_{\text{Cp}}=\frac{d_{s}}{T_{s}}\left\vert \frac{dT_{c}}{dd_{n}}\right\vert
=d_{s}\left\vert \frac{d\left(  \ln\left(  T_{c}\right)  \right)  }{dd_{n}%
}\right\vert =\frac{N_{n}}{N_{s}}\frac{\left(  NV\right)  _{s}-\left(
NV\right)  _{n}}{\left(  \left(  NV\right)  _{s}\right)  ^{2}} \label{S_Cp}%
\end{equation}

The predictions of equation (\ref{S_Cp}) for the normalized initial slope
(which is derived from equation (\ref{Tc_Cp})) has two problems, (i) Pb is not
a superconductor\ with $T_{c}<<\Theta_{D}$ and therefore the equation
(\ref{Tc_Cp}) does not yield a good representation of Pb, (ii) equation
(\ref{Tc_Cp}) uses the same Debye temperature for both metals and this is
generally not fulfilled in the experiment. One can replace the BCS-Cooper
formula for $T_{c}$ by an (implicit) expression
\begin{equation}
\frac{1}{\left(  VN\right)  _{ef}}=\sum_{n=0}^{n_{c}}\frac{1}{\left(
n+\frac{1}{2}\right)  } \label{Tc_G}%
\end{equation}
where $n_{c}=\Theta_{D}/\left(  2\pi T_{c}\right)  $. We do not evaluate
equation (\ref{Tc_G}) here because its results are in included in our
numerical procedure as a limiting case.

Werthamer \cite{W32} derived a set of equations for the transition temperature
of a double layer of a super- and a normal conductor in the dirty limit, using
de Gennes' interface boundary condition \cite{D12}. In this theory the gap
parameter is approximated as a cosine and hyperbolic cosine function in the
superconductor and normal conductor respectively. (This is sometimes called
the single mode expansion). This theory agrees quite well with the
experimental results for double layers of two superconductors \cite{B133}. The
normalized initial slope can be expressed as%

\begin{equation}
S_{\text{Wh}}=\frac{N_{n}}{N_{s}}\frac{\pi^{2}}{4}\chi^{-1}(-\ln(\frac{T_{s}%
}{T_{n}}))
\end{equation}
here $\chi^{-1}\left(  y\right)  $ is the inverse function of $\chi
(x)=\Psi(\frac{1}{2}+\frac{1}{2}x)-\Psi(\frac{1}{2})$ and $\Psi(z)$ is the
digamma function. If we assume that the transition temperature $T_{n}$ for the
normal metals Cu, Ag, Au and Mg is infinitely small, then $\chi^{-1}%
(-\ln(\frac{T_{s}}{T_{n}}))$ takes the value one. The values of $S_{\text{Wh}%
}$ according to Werthamer's theory are included in table I. The values of
$S_{\text{Wh}}$ don't show a good agreement for the normal metals Cu, Ag, Au
and Mg.

Finally we compare the experimental values $S_{\text{exp}}$ with our numerical
calculation. This numerical calculation derives the transition temperature of
a double or multi layer of a superconductor and a normal conductor. The
equivalence in the propagation of the superconducting pair amplitude and a
single electron in Gorkov's linear gap equation is used. The single electrons
act as messengers which carry the information about the superconducting gap
($N_{s}\Delta\left(  \mathbf{r}^{\prime}\right)  /\tau_{T}$) from one
position-time $\left(  \mathbf{r}^{\prime},t^{\prime}<0\right)  $ to another
position-time $\left(  \mathbf{r},t=0\right)  $. This message which decays
thermally with time as $\eta_{T}\left(  t\right)  =\sum_{\left\vert \omega
_{n}\right\vert <\Omega_{D}}\exp\left(  -2\left\vert \omega_{n}\right\vert
\left\vert t^{\prime}\right\vert \right)  $, is integrated at $\left(
\mathbf{r,}t=0\right)  $ over all starting position-times $\left(
\mathbf{r}^{\prime},t^{\prime}\right)  $ and, after multiplication with the
BCS interaction $V_{s},$ \ yields the new gap function $\Delta\left(
\mathbf{r}\right)  $. At the transition temperature the procedure has to be
self-consistent, i.e. the initial and final gap function have to be identical.
The propagation of the single electrons is then quasi-classically simulated.
The frame work of the calculation is the weak coupling theory of superconductivity.

This numerical procedure to calculate the transition temperature of double or
multi-layers consisting of thin films of superconductors and normal conductors
is very flexible. It uses the following input parameters of the individual
metal films (i) thickness, (ii) density of states and Fermi velocity, (iii)
transition temperature, (iv) Debye temperature, (v) mean free path and (vi)
transmission through the interface between the films.

An important outcome of the numerical simulation is the result that the
normalized initial slope of an SN double layer as a function of $d_{n}$ at
$d_{n}=0$ does not depend on (i) the mean free paths of the two metals, (ii)
the thickness of the superconductor and (iii) a finite (but not too large)
barrier between the two metals.

We include the numerical results $S_{\text{sim}}$ in table I. For the double
layers with the normal conducting metals Cu, Ag, Au and Mg the discrepancy
between experiment and theory is very large, of the order of 3. It is
remarkable that the deviation is considerably smaller when the
\textquotedblright normal metal\textquotedblright\ is really a superconductor
with a smaller transition temperature although the deviation is still factor
of about 1.5. For In and Sn, however, there is a good agreement between theory
and experiment.

It is surprising that this disagreement between experiment and theory has not
been noticed in the past. The main reason is that the majority of the
experimental and the theoretical work focused on NS sandwiches with thick
normal metal films. Then superconductivity is only obtained for a finite
thickness of the superconductor. In this case a comparison between experiment
and theory requires many fit parameters such as the transparency of the
interface and the mean free paths of the superconductor and the normal
conductor. Therefore it is quite possible to fit the experimental data by
using the wrong parameters that can't be checked otherwise.

The physical origin of this disagreement between experiment and theory is not
understood. Our theoretical simulation of the SPE uses the frame work of weak
coupling superconductivity. Quench condensed Pb, In and Sn are not weak
coupling. The ratios of $2\Delta_{0}/\left(  k_{B}T_{s}\right)  $ for quench
condensed films are 4.6 for Pb, 3.9 for In and 4.0 for Sn \cite{B6},
\cite{B75}. An obvious proposal would be to solve the superconducting
proximity effect for strong coupling superconductors. This means to develop
and solve a series of equations for the energy and position dependent gap
function $\Delta\left(  \mathbf{r},\omega_{l}\right)  $. This would be an
extremely demanding job. As a start, we considered the thin film limit for the
SPE of strong superconductors. This consideration, which will be published
elsewhere, does not remove the discrepancy between experiment and theory.

The fact that the Werthamer result $S_{\text{Wh}}$ disagrees less with the
experimental data $S_{\text{exp}}$ for Cu, Ag, Au and Mg than our numerical
result might be accidental. The single mode expansion of $\Delta\left(
\mathbf{r}\right)  $ in the Werthamer theory is surely less appropriate than a
self-consistant gap function.

The experimental normalized initial slope is proportional to the density of
states ratio. Although the density of states can be modified in quench
condensed films it is inconceivable that this explains a factor of three in
the initial slope. Let us return to Fig.1 where the reduction of $T_{c}$ of Pb
by a thin layer of Ag is plotted.

\section{Conclusion}

In this paper, the superconducting proximity effect is investigated for SN
double layers in a regime where the resulting $T_{c}$ does not depend on the
mean free path of the films and, within limits, not on the transparency of the
interface. This regime includes the thin film limit and the normalized initial
slope $S_{sn}=(d_{s}/T_{s})|dT_{c}/dd_{n}|$ . While the layer S is always Pb
the layer N is either a non-superconducting metal such as Cu, Ag, Au and Mg or
a superconductor with a $T_{c}$ below the transition temperature of Pb. The
experimental results for the transition temperature $T_{c}$ are compared with
a numerical calculation which was recently developed in our group. The results
for the SN double layers can be divided into three groups:

\begin{itemize}
\item When N represents a non-superconducting metal film (N=Cu, Ag, Au and Mg)
we observe grave deviations between experiment and theory by a factor of the
order of three.

\item When N represents a superconductor with a low $T_{c}$ (N=Cd, Zn, Al) the
deviation between experiment and theory is still there but reduced by a factor
of two.

\item When N represents a superconductor with a $T_{c}$ which is about half
the $T_{c}$ of Pb (N=In, Sn) then we observe a reasonably good agreement
between experiment and theory.
\end{itemize}

Prior to our recent experiments we believed that the proximity effect between
a superconductor and a normal conductor represented an intensively studied
phenomenon with a good theoretical understanding. We are deeply puzzled by the
large observed discrepancy between experiment and theory. It would be very
desirable if other theoretical approaches would give quantitative predictions
for the normalized initial slope in SN double layers. There have been a number
of theoretical papers published which extended the proximity effect to more
complex systems, for example between a superconductor and a ferromagnet but
which include implicitly the simpler case of an SN double layer. These authors
would be able to calculate quantitatively the normalized initial slope from
their theory.

Experimentally it would be desirable to extend the measurements to SN layers
where S is a weak coupling superconductor. This requires lower temperatures
but permits the use of thicker films because the coherence lengths are larger
at lower temperatures. Aluminum would be a good candidate for the
superconductor if evaporated in ultra-high vacuum so that the Al is not
granular.

\end{document}